# Recent Progress on Developments and Characterization of Hybrid CMOS X-ray Detectors

A. D. Falcone[a], Z. Prieskorn[a], C. Griffith[a], S. Bongiorno[a], D. N. Burrows[a]

[a]Penn State University, Dept. of Astronomy & Astrophysics, University Park, PA 16802 USA;

## ABSTRACT

Future space-based X-ray telescope missions are likely to have significantly increased demands on detector read out rates due to increased collection area, and there will be a desire to minimize radiation damage in the interests of maintaining spectral resolution. While CCDs have met the requirements of past missions, active pixel sensors are likely to be a standard choice for some future missions due to their inherent radiation hardness and fast, flexible read-out architecture. One form of active pixel sensor is the hybrid CMOS sensor. In a joint program of Penn State University and Teledyne Imaging Sensors, hybrid CMOS sensors have been developed for use as X-ray detectors. Results of this development effort and tests of fabricated detectors will be presented, along with potential applications for future missions.

## 1. INTRODUCTION

While Chandra, XMM, Swift, Suzaku, and other X-ray telescopes continue to operate and a limited number of small specialized X-ray missions (e.g. GEMS) are planned, it is clear that X-ray astronomy is in dire need of a new, large-area and high-resolution observatory to achieve the science goals of the coming decades. There may be multiple medium-to-large and specialized observatories, or there may be one large, general-purpose observatory (e.g. SMART-X or Gen-X); the future program is still not firmly defined. However, it is already clear that upcoming missions intend to use high throughput (e.g. collecting area ~10x that of XMM and Chandra) to achieve their scientific requirements, and some missions intend to have high spatial resolution [1]. This high throughput is largely driven by the need for enhanced spectral resolution, which in turn, drives a need for improved photon counting statistics. This need for higher throughput leads to a need for focal plane detectors with improved capabilities, since the previous generation of CCD detectors would suffer from saturation effects (pile-up) and radiation damage effects if they were placed in the focal plane of a mission such as SMART-X. One new technology that could be used to satisfy this need is that of hybrid CMOS detectors (a type of active pixel sensor), which offer addressable pixel readout (independent of all other pixels on detector), >10 MHz readout rates for each parallel readout line, inherent radiation hardness, and very low power needs.

## 2. MOTIVATION FOR CMOS

### 2.1 Current state of the art

CCDs continue to be today's state-of-the-art X-ray detectors. This is primarily due to their large detector format, high spatial resolution, good quantum efficiency, and nearly Fano-limited energy resolution. However, CCDs have a number of limitations that become especially serious for long-lived, high-throughput X-ray missions. The most important of these are pile-up limitations, radiation damage associated problems, and their high power requirements. Current X-ray missions are already severely limited by pile-up (i.e. saturation), which is caused by high photon count rates leading to multiple X-ray events in a single pixel. This problem will become much more severe with the high throughput missions currently in the planning stages, and will effect both energy resolution (due to the total charge from multiple photons being summed and misconstrued as the charge from one photon) and flux measurements (due to the undercounting of photons when pile-up is present). Additionally, radiation damage severely limits the operating lifetime of current X-ray instruments, decreasing the energy resolution with time due to proton displacement damage in the silicon lattice.

Detectors for new missions will need to overcome these limitations if they are to achieve the high throughput and energy resolution that is dictated by the science requirements.

## 2.2 Advantages of Hybrid CMOS detectors

The X-ray astronomy community can benefit from past developments that have already led to the maturity of hybrid CMOS detectors (HCDs) for optical and infrared applications [2,3,4]. These prior developments have brought the technology readiness level of the devices and their readout ASICS to a high level, thus bringing us to the point where relatively small modifications are all that is required to enable their use in future X-ray missions. Existing devices developed by Teledyne Imaging Sensors (TIS) have backside illumination with stable backside surfaces and negligible dead layers, large detector formats (4096 × 4096), moderate read noise, and much higher radiation hardness than any CCD detector can achieve [4]. Furthermore, they have random-access pixel readout, which allows much more flexible readout schemes than CCDs can accommodate. In comparison with CCDs, HCDs have the following advantages:

Pileup: because HCDs have randomly addressable pixel readout, only the pixels containing interesting sources need to be read out. An HCD can do a high speed read out to find pixels with X-rays, and a slow (or multiple rapid) read out of those pixels to get the best noise performance. The HCD need only do this for a small window containing the source of interest, while a CCD must read all pixels. Existing HCDs can achieve readout times for small target windows as low as 30 μs, allowing very high count rate applications with minimal pileup (> 10,000 counts/s for typical X-ray telescope designs), while also providing excellent timing information. This can lead to two to three orders of magnitude improvement in peak flux capability relative to CCDs.

Radiation Damage: The CMOS structures of the readout multiplexer are inherently radiation hard to levels far in excess of those required for any astronomical mission (> 100 krads). The devices feature direct readout of every pixel, thereby avoiding the charge transfer problems of CCDs, and essentially eliminating sensitivity to proton displacement damage (charge must only be transferred through the ~100μm thickness of the Si absorber array, rather than across its ~few cm width). The result is a device that is orders of magnitude less sensitive to radiation damage than CCD detectors, without resorting to techniques such as charge injection.

Micrometeoroids: *XMM-Newton* and *Suzaku* have lost all or parts of CCDs on-orbit, probably due to micrometeoroid damage to their gate structures. The *Swift* XRT has lost the use of several columns due to micrometeoroid-damaged pixels causing high dark current that blooms up the columns. The lack of exposed gates protects HCDs from the first failure mechanism, and the direct pixel readout prevents individual pixels from blooming across the detector. We therefore expect HCDs to be more robust against micrometeoroid damage than CCDs.

Readout noise: readout noise in these devices is currently higher than that achieved in the best CCDs at slow readout speeds, but is better than CCDs at Megapixel/s speeds [4]. In the future, it is possible that read noise as low as 1-2 $e^-$ can be achieved in CMOS readout circuitry [5]. Since the CMOS readout is non-destructive, multiple reads of X-ray signals can be averaged to reduce noise levels even further, as has been demonstrated for CCDs in the past [6].

Low Power: on-board integration of camera drive electronics and detector signal processing reduces power consumption and mass in comparison with traditional CCD camera designs [7], and results in a more reliable instrument for space applications. For example, the Swift-XRT requires a total of 8.4 W to produce and drive the CCD readout signals, while this function can now be achieved, even faster, with <100 mW of power using a H1RG HCD with a SIDECAR ASIC.

## 3. DETECTOR DESIGNS AND MODIFICATIONS

### 3.1 The Detectors

The detectors chosen for evaluation in this project were modified versions of the Teledyne Hawaii-1RG (H1RG) hybrid CMOS detectors, as well as one specially modified H2RG hybrid CMOS detector. The HxRGs were initially developed for optical/IR astronomy purposes, and their development is at an advanced stage with high TRL for the optical/IR devices. There are plans for flight of an H2RG 2048x2048 pixel HCD on the James Webb Space Telescope. The JWST devices use a different detector material optimized for optical/IR (HgCdTe), rather than the Si used in the HyViSI devices needed for X-ray detection, but the HxRG read out integrated circuit (ROIC) is the same. Optical-optimized HyViSI devices have reached high TRL levels through the Mars Reconnaissance Orbiter and the Orbiting Carbon Observatory programs, but these devices are not optimized for X-rays [14]. To perform the initial evaluation of these detectors as X-ray devices, the first step was to make a device with the anti-reflection coating removed, thus enhancing the quantum efficiency in the X-ray-UV range, and to apply an aluminum optical blocking layer. These initial detectors were made with 1024x1024 pixels with 18 μm pitch (i.e. standard H1RG parameters) with the standard HxRG source follower readout structure in the ROIC. Another modified H1RG detector, which will be reported in a separate publication, was made with 36 μm pitch pixels with four individual readout lines attached to each pixel, using the 18 μm spacing of the standard H1RG ROIC. A second batch of two detectors was made with changes aimed at achieving a reduction in dark current. This second batch also experimented with changes to the optical blocking filter thickness, as described below. Lastly, a device with an H2RG ROIC with one out of every 4 readout lines bump bonded to a 36 μm pixel pitch detector array, resulting in wider spacing between ROIC lines, was tested in order to evaluate the impact of this change on inter-pixel coupling that causes charge from one pixel to be observable in surrounding pixels (see [12]).

### 3.2 The Aluminum blocking filter

Optical light blocking filters are typically required on X-ray telescopes due to the fact that optical light from stars and other bright background sources will contaminate the X-ray detector images, which are taken in a single photon counting mode. Typical filter thicknesses range from several hundred Angstroms to ~1500 Angstroms. These filters are typically applied to substrates that are then mounted in front of the detector. While the filter is an important requirement for any X-ray telescope detector (either CCD or hybrid CMOS), an unfortunate side effect is that the filter and the substrate on which the filter is applied change the response of the entire system, particularly at the low range of the sensitive energy band where several scientifically important X-ray lines reside. This effect is present in modern observatories, such as Chandra and XMM, and it must always be accounted for when calculating the response of the instruments. By eliminating the substrate and applying the filter directly to the detector, one can eliminate much of this impact on the instrument response.

The Aluminum blocking filter for the detectors described in this paper has been directly deposited on the hybrid CMOS detector. Figure 1 shows an image of one of the test detectors that was made with the Al blocking filter covering half of the active area of the detector, while the other half was left free of any filters. This format has allowed us to test the performance of the detector both with and without the filter to be certain there were no adverse effects caused by the direct deposition. This also allowed us to measure the absorption of light by the filter by comparing the response of the two halves. For the detectors described in this report, the Al optical blocking layer ranged in thickness from 180 to 1000 Angstroms. The right side of Figure 1 shows some transmission curves for these filters.

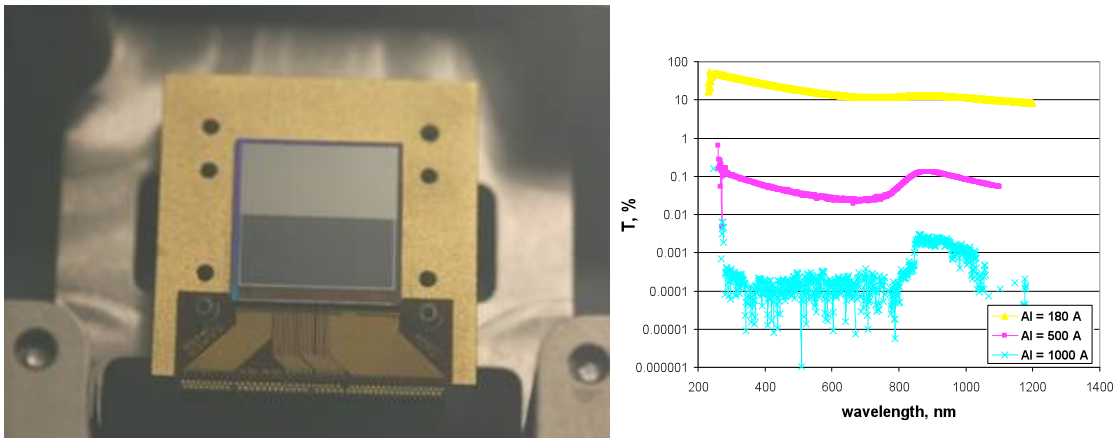

Figure 1: (Left) Photo of one of the test devices, H1RG-118. The aluminum blocking filter is the light gray area covering the top half of the detector array. Half of the detector was left with no aluminum filter for testing purposes. (Right) Optical transmission curves for 3 HCD Al optical blocking filters from our program.

# 4. OUR TEST STAND

Two test stands have been used at PSU to characterize the operation and properties of these detectors, one with a SIDECAR ASIC (Testcam 2) and one with a custom headboard and modified ADC and sampling electronics boards used for prior testing of Swift and other CCDs (Testcam 1).

## 4.1 Testcam 1

A custom headboard was designed to interface with the standard output, programming, bias voltage, and clocking lines of the H1RG detector package. In addition to a low noise, multi-layer design for the electronic layout and the signal amplification, this circuit board incorporated a rigid ring that is abutted to a vacuum chamber on both sides of the board. This headboard and a mounted detector can be seen in Figure 2. This architecture allows us to sandwich the board between two halves of a vacuum chamber and carry the signal lines out through the internal layers of the board without the need for many bulk head connections on the chamber. It also allows us to place signal conditioning and amplification electronics very close to the initial output from the detector, while simultaneously allowing us to probe lines and change components without breaking the seal on the vacuum, which achieved pressures as low as $\sim 2 \times 10^{-7}$ mbar. The headboard was also coated with black solder mask to reduce the risk of light leaks. The downstream signal conditioning and bias and clock generation were accomplished using modified CCD readout and clocking boards that were previously used for testing of the Swift CCDs. The detector package itself was mounted to a copper cold finger that allowed us to cool the detector with liquid Nitrogen. The detector could be illuminated with an Fe-55 X-ray source that was mounted on one side of the chamber behind a controllable Aluminum shield, and it could be illuminated with switchable LEDs mounted in the chamber.

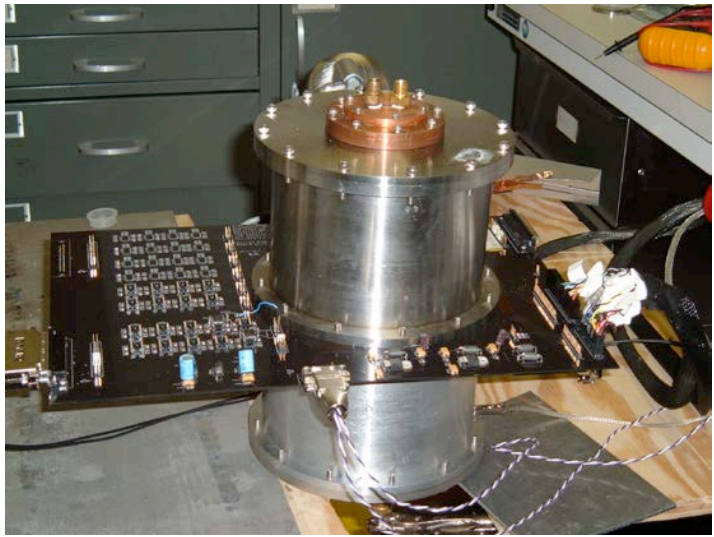 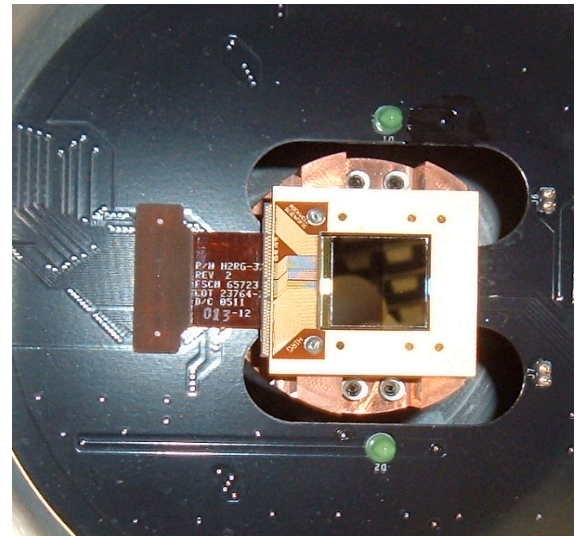

Fig. 2. [left] Laboratory Test Camera for Hybrid CMOS detectors (shielding enclosure has been removed from the head board for illustrative purposes). [right] Hybrid CMOS detector mounted on PSU headboard for testing. The HCD is flanked by test LEDs.

## 4.2 Testcam 2

For the second generation detectors, we developed TestCam2 (Figure 3) in order to measure energy resolution as a function of incident X- ray energy and in order to test the readout of the detector using a high TRL ASIC that would be the likely choice for most space missions. This camera uses a Teledyne SIDECAR ASIC [8] to drive the detector and to digitize the analog signals. The X-ray can be produced with an Fe-55 X-ray source or with an alpha source that has multiple X-ray florescence targets available on a filter wheel, thus allowing us to produce X-rays of various energies. In this camera, the detector is cooled with the same liquid nitrogen cold finger, while the SIDECAR ASIC is uncooled in the vacuum chamber.

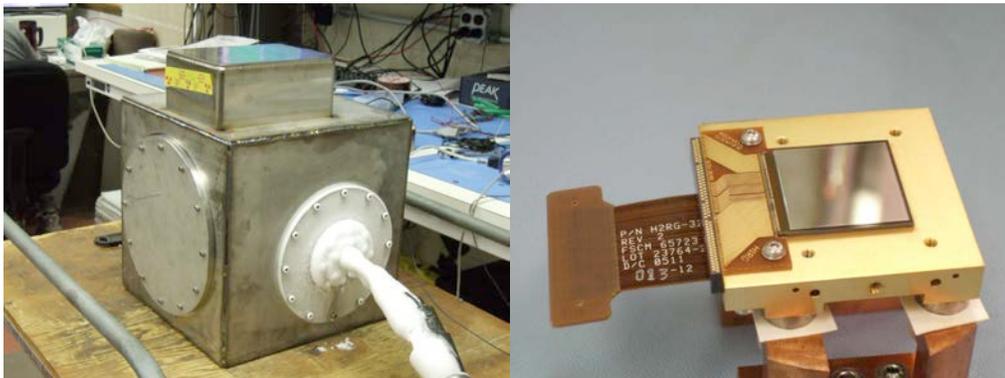

Figure 3: (left) Testcam 2 with internally mounted SIDECAR ASIC, HxRG detector, X-ray sources (not visible in picture), and cold finger being chilled with liquid notrogen. (Right) Photo of H1RG-161, which has a complete 100 nm Al optical blocking filter, ready to be mounted to cold finger of Testcam 2.

## 5. TEST RESULTS

An overview of the measurements for each of the detectors is shown in Table 1. The read noise measurements were obtained by taking 100 dark images with each detector and making a histogram of the resulting measured charge in each pixel. The histograms were well fit with a Gaussian distribution, and the average RMS value is reported. For H1RG-125, we report two separate values for the read noise (and the ΔE/E) since this detector was damaged by being exposed to a bright light source while having an applied substrate voltage. This damage increased the noise in the detector (see [10] for more detailed description) so we report the pre-damage and post-damage values.

For measurements of detector dark current, we used 3 ramps comprised of 400 frames per ramp, with 5.2 sec exposures for each frame. In this way, the dark current charge accumulates frame-by-frame in each pixel, allowing us to measure the dark current by measuring the slope of the charge versus time plot. We report the average dark current over the detector. The logarithm of dark current versus temperature is more steeply sloped at temperatures above approximately 180 K, following a relationship approximately proportional to $T^{1.5}e^{-(E/2kT)}$, where E is the silicon band gap energy. If one extrapolates these higher temperature dark current measurements down to 150 K, then a much lower value is obtained for the dark current, relative to our measurement at 150 K. Between ~150 K and 180 K, the dark current follows a much shallower slope, which is probably dominated by dark current from surface interfaces, as opposed to the bulk silicon that probably dominates at higher temperatures.

As initially reported in Falcone et al. 2007 [9] and detailed in Bongiorno et al. 2009 & 2010 [10,11], the optical blocking filters have worked as expected and the detectors are efficient detectors of X-rays, with the caveat that the early generations have been plagued by interpixel capacitive crosstalk (IPC). This effect causes some of the charge in a pixel to be measurable in neighboring pixels, and has the primary effect of degrading energy resolution. The measurement of the IPC and the ΔE/E is described in Griffith et al. 2012 [12; these proceedings], and the early measurements are described in Bongiorno et al. 2010. This IPC can be minimized by using larger separations between ROIC connections, as can be seen by comparing the IPC measured for H2RG-122 (≤1.7%), which has ROIC bond connections every 36 μm, to the ~6% IPC for the other detectors, which have 18 μm spacing. However, since H2RG-122 is an engineering grade detector with higher read noise and variable response across the detector, this IPC improvement was not enough to provide a significantly improved value for the energy resolution, ΔE/E. The next generation of these detectors will use improved CTIA amplifiers in the ROIC, which will eliminate any measurable IPC, even for small pixel and ROIC bond spacing. These developments have already begun.

Most of the results presented in this paper, and in particular in Table 1, were obtained with Testcam 2, using the SIDECAR ASIC, but it is worth mentioning that our Testcam 1 results are consistent with those presented here, whenever comparison is possible. In particular, it can be seen from Figure 4, that X-ray events from the 1st generation of HCDs are typically blurred as non single pixel events, displaying the IPC discussed above. While completely expected, this provide confirmation that this effect is in the H1RG ROIC; not the SIDECAR.

Further characterization of these detectors and the next generation of HCDs is ongoing. Final results will be reported in a forthcoming publication.

Table 1 shows a summary of our test results to-date.

| | Al Filter thickness (Å) | RMS Read Noise ($e^-$) | Dark $I$ 150 K, Data[a] (e-/s/pix) | Dark $I$ 293 K, Fit[b] (e-/s/pix) | $\Delta E/E$ @ 5.9 keV (FWHM) | $\Delta E/E$ @ 1.5 keV (FWHM) | IPC (avg. crosstalk to neighbor pixel) |
|---|---|---|---|---|---|---|---|
| **H1RG-125 (before damage)[c]** | 500 (half) | 7.5 | | | 0.042 | | |
| **H1RG-125 (after damage)[c]** | 500 (half) | 10.7 | 0.214 ± 0.025 | 3.78E6 ± 6.27E4 | 0.062 | 0.20 | 6.4% |
| **H1RG-161** | 1000 | 10.8 | 0.020 ± 0.005 | 8.15E5 ± 2.83E4 | 0.092 | 0.16 | 6.1% |
| **H1RG-167** | 180 (half) | 7.0 | 0.056 ± 0.026 | 4.45E6 ± 7.89E5 | 0.050 | 0.17 | 6.1% |
| **H2RG-122** | none | 17.3 | 0.020 ± 0.001 | 9.38E6 ± 8.09E5 | 0.075 | 0.18 | ≤1.7% |

[a]The directly measured dark current values at 150 K.
[b]These dark current values are from an extrapolation of a fit to the dark current data from 180 K to 210 K.
[c]H1RG-125 was damaged during early measurements, when it was exposed to light while under a bias voltage of ~18 V. The pre-damage values (reported in Bongiorno et al. 2009 [10]) are shown here, as well as post-damage measurements.

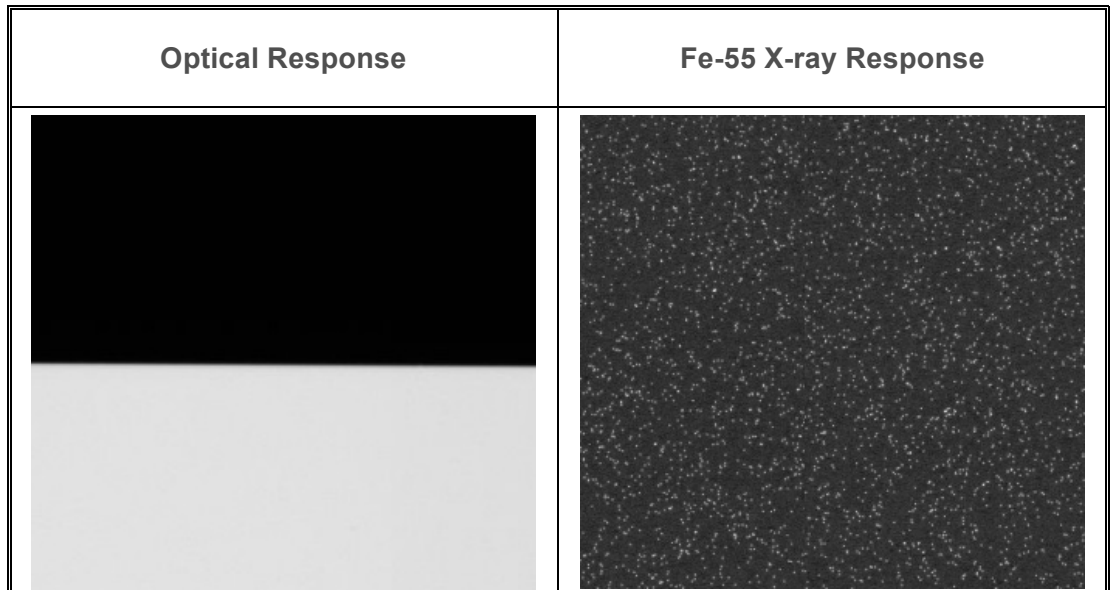

Figure 4. Optical (left) and X-ray (right) images from an H1RG hybrid CMOS detector with 1024x1024 18um pixels. These images were taken with Testcam 1. The Al filter covering the top half of the device has low optical transmission. Each speckle in the X-ray image is an X-ray from an Fe-55 source. The image uses a 10 sec integration and a single correlated-double-sample subtraction.

## 6. DISCUSSION AND CONCLUSIONS

The replacement of CCDs with hybrid CMOS detectors for X-ray astronomy is motivated by the need for the radiation hardness, fast readout speeds, low power, and flexible readouts that hybrid CMOS detectors provide. We have tested several engineering grade H1RG and H2RG X-ray hybrid CMOS detectors. We found read noise as low as 7.0 $e^-$ RMS, dark current as low as $1.8\times10^{-4}$ $e^-$ $s^{-1}$ $pixel^{-1}$ (corresponding to ~40 nA $cm^{-2}$) extrapolated to 150 K, and $\Delta E/E$ as low as 4.2% at 5.9 keV, with typical IPC values of ~6%. We also found that the IPC could be reduced to below 1.7% by using 36 μm spacing between the ROIC bump bonds. This performance is suitable for some specific X-ray instruments (e.g. JANUS [13]), but to make these detectors viable for future large missions such as SMART-X, the best of each of these parameters will be needed, and low IPC will be needed, even for smaller pixel sizes. This should be achievable by taking the best aspects of these detectors and combining them with amplifiers that are free of IPC problems. These developments are forthcoming.

## 7. ACKNOWLEDGEMENTS

We gratefully acknowledge Teledyne Imaging Systems, particularly James Beletic and Yibin Bai, for providing useful comments and for loaning us the modified H2RG detector. This work was supported by NASA grants NNG05WC10G, NNX08AI64G, and NNX11AF98G.